\documentclass[twocolumn,showpacs,floatfix,
superscriptaddress,
longbiblography,
amsmath,amssymb,
aps,
prl]{revtex4-2}

\usepackage{graphicx}
\usepackage{dcolumn}
\usepackage{amsmath}
\usepackage{bm}
\usepackage{appendix}
\usepackage{soul}
\usepackage{multirow}
\usepackage[bookmarks=true,colorlinks=true,urlcolor=blue,linkcolor=blue,citecolor=blue,breaklinks]{hyperref}
\usepackage[mathlines]{lineno}

\usepackage{inconsolata} 
\usepackage[usenames,dvipsnames]{color} 
\usepackage{listings}

\usepackage{pdfpages} 
\usepackage{pgffor} 
\usepackage{xr} 

\makeatletter
\AtBeginDocument{\let\LS@rot\@undefined}
\makeatother

\newcommand{\jm}[1]{{\color{OliveGreen} #1} }

\renewcommand{\selectlanguage}[1]{}
\begin{document}

\preprint{APS/123-QED}

\title{Critical Dynamics of the Anderson Transition on Small-World Graphs}


\author{Weitao Chen}
\affiliation{Department of Physics, National University of Singapore, Singapore.}
\affiliation{MajuLab, CNRS-UCA-SU-NUS-NTU International Joint Research Unit, Singapore.}
\affiliation{Centre for Quantum Technologies, National University of Singapore, Singapore.}
\author{Ignacio García-Mata}
\affiliation{Instituto de Investigaciones Físicas de Mar del Plata (IFIMAR),
CONICET–UNMdP, Funes 3350, B7602AYL Mar del Plata, Argentina.}
\affiliation{Consejo Nacional de Investigaciones Científicas y Tecnológicas (CONICET), Argentina.}
\author{John Martin}
\affiliation{Institut de Physique Nucléaire, Atomique et de Spectroscopie,
CESAM, University of Liège, B-4000 Liège, Belgium.}
\author{Jiangbin Gong}
\affiliation{Department of Physics, National University of Singapore, Singapore.}
\affiliation{MajuLab, CNRS-UCA-SU-NUS-NTU International Joint Research Unit, Singapore.}
\affiliation{Centre for Quantum Technologies, National University of Singapore, Singapore.}
\author{Bertrand Georgeot}
\affiliation{Laboratoire de Physique Théorique, Université de Toulouse, CNRS, UPS, France.}
\author{Gabriel Lemarié}%
\email{ lemarie@irsamc.ups-tlse.fr}
\affiliation{MajuLab, CNRS-UCA-SU-NUS-NTU International Joint Research Unit, Singapore.}
\affiliation{Centre for Quantum Technologies, National University of Singapore, Singapore.}
\affiliation{Department of Physics, National University of Singapore, Singapore.}
\affiliation{Laboratoire de Physique Théorique, Université de Toulouse, CNRS, UPS, France.}

\date{\today}

\begin{abstract}
The Anderson transition on random graphs draws interest through its resemblance to the many-body localization (MBL) transition with similarly debated properties. In this Letter, we construct a unitary Anderson model on Small-World graphs to characterize long time and large size wave-packet dynamics across the Anderson transition. We reveal the logarithmically slow non-ergodic dynamics in the critical regime, confirming recent random matrix predictions. Our data clearly indicate two localization times: an average localization time that diverges, while the typical one saturates. In the delocalized regime, the dynamics are initially non-ergodic but cross over to ergodic diffusion at long times and large distances. Finite-time scaling then allows us to characterize the critical dynamical properties: the logarithm of the average localization time diverges algebraically, while the ergodic time diverges exponentially. Our results could be used to clarify the dynamical properties of MBL and could guide future experiments with quantum simulators.


\end{abstract}

\maketitle

\emph{Introduction.\textemdash} The Anderson transition is a well-known metal-insulator transition driven by the interplay of interference and disorder \cite{PhysRev.109.1492, abrahams201050, RevModPhys.57.287, Kramer_1993, RevModPhys.80.1355}. While it is well understood in low-dimensional systems—both theoretically \cite{PhysRevB.19.783, Efetov1983, WEGNER1989663, RevModPhys.80.1355}, numerically \cite{PhysRevLett.76.1687,PhysRevB.65.113109,PhysRevB.75.174203,doi:10.7566/JPSJ.83.084711,PhysRevLett.82.382,PhysRevB.84.134209,PhysRevLett.105.046403,PhysRevLett.102.106406,PhysRevLett.105.090601}, and experimentally \cite{Garreau2017,PhysRevLett.103.155703,PhysRevLett.101.255702,PhysRevLett.108.095701,PhysRevLett.45.1723,PhysRevLett.48.1284,PhysRevB.98.214201,hu2008localization,jendrzejewski2012three,semeghini2015measurement}—its critical properties in the high-dimensional limit remain a subject of debate  \cite{PhysRevB.95.094204, zirnbauer2023wegner, altshuler2024renormal}.

Interest has naturally turned to the study of the Anderson transition in infinite dimensions ($\text{AT}^{\infty}$) \cite{PhysRevLett.72.526, mirlin1994statistical, arenz2023wegnermodeltreegraph, zirnbauer2023wegner, PhysRevB.109.174216}. Early efforts focused on the Anderson model on the Bethe lattice \cite{PhysRevB.34.6394, ZIRNBAUER1986375, Monthus_2011, biroli2012difference, Parisi_2020, PhysRevB.105.094202, PhysRevLett.113.046806, PhysRevB.96.214204, KRAVTSOV2018148}. Recently, strong interest has arisen in the Anderson model defined on random graphs featuring effective infinite dimensionality, such as Random Regular Graphs (RRG) and Small-World Graphs (SWG), due to their resemblance to the Fock-space graph of a Many-Body Localized (MBL) system \cite{PhysRevLett.78.2803, PhysRevLett.95.206603, BASKO20061126, TIKHONOV2021168525}. The MBL transition is another ergodic-to-non-ergodic phase transition for isolated quantum many-body systems driven by disorder, with crucial implications for the foundations of statistical mechanics and quantum information theory \cite{RevModPhys.91.021001, sierant2024manybody,collins2016random}. Several key insights from the Anderson model on random graphs have found successful applications in advancing the understanding of the MBL problem and vice versa \cite{PhysRevB.96.201114, PhysRevLett.118.166801, PhysRevResearch.2.012020, PhysRevB.106.214202, PhysRevLett.123.180601, PhysRevLett.125.250402, TIKHONOV2021168525, PhysRevB.101.134202, PhysRevResearch.2.023159, PhysRevB.103.064204, PhysRevB.104.024202, PhysRevB.104.174201, PhysRevB.106.054203, PhysRevB.109.214203, PhysRevB.110.014205, sun2024characterizing, niedda2024renormal}, providing valuable analogies.

The two transitions have been marked by ambiguity and challenges: the existence or not of a non-ergodic delocalized phase and the critical properties of the Anderson transition \cite{PhysRevB.94.184203,PhysRevB.109.174216, ADMirlin_1991, fyodorov1992novel,PhysRevLett.76.1687,PhysRevB.99.024202,PhysRevB.99.214202,PhysRevB.110.184210,PhysRevResearch.6.L032024,PhysRevB.110.014210,Kravtsov_2015,kravtsov2020localization,PhysRevResearch.2.043346,PhysRevResearch.2.012020,10.21468/SciPostPhys.6.1.014,10.21468/SciPostPhys.11.2.045, PhysRevB.94.220203,PhysRevLett.118.166801,PhysRevLett.117.156601,KRAVTSOV2018148,SciPostPhys.15.2.045,PhysRevResearch.2.042031,PhysRevB.109.184204, PhysRevB.106.214202, doi:10.1073/pnas.2401955121}, as well as the value of the critical disorder and the nature of the transition for the many-body localized problem \cite{PhysRevB.95.155129,PhysRevB.105.174205,PhysRevB.106.L020202,leonard2022signatures,PhysRevB.100.104204,PhysRevE.102.062144,PhysRevB.102.064207,PhysRevB.103.024203,PhysRevE.104.054105,PhysRevLett.127.230603,ABANIN2021168415,PhysRevLett.124.186601,Panda_2019,PhysRevB.102.100202,PhysRevLett.133.116502}. Among the most debated aspects are the critical dynamics. These systems commonly exhibit slow relaxation, strongly affected by finite-size effects \cite{Panda_2019,ABANIN2021168415, TIKHONOV2021168525}, which have fueled controversies regarding the nature of the delocalized phase of the AT$^\infty$ \cite{PhysRevB.98.134205,PhysRevB.101.100201,PhysRevB.105.174207} and even the very existence of an MBL phase \cite{PhysRevLett.131.106301,PhysRevB.100.104204,PhysRevB.103.024203, PhysRevB.108.134204,PhysRevB.109.224206,PhysRevB.108.L140201}.
This highlights the fundamental importance of these problems within the broader context of phase transition research.

\begin{figure*}
\includegraphics[width=0.99\textwidth]{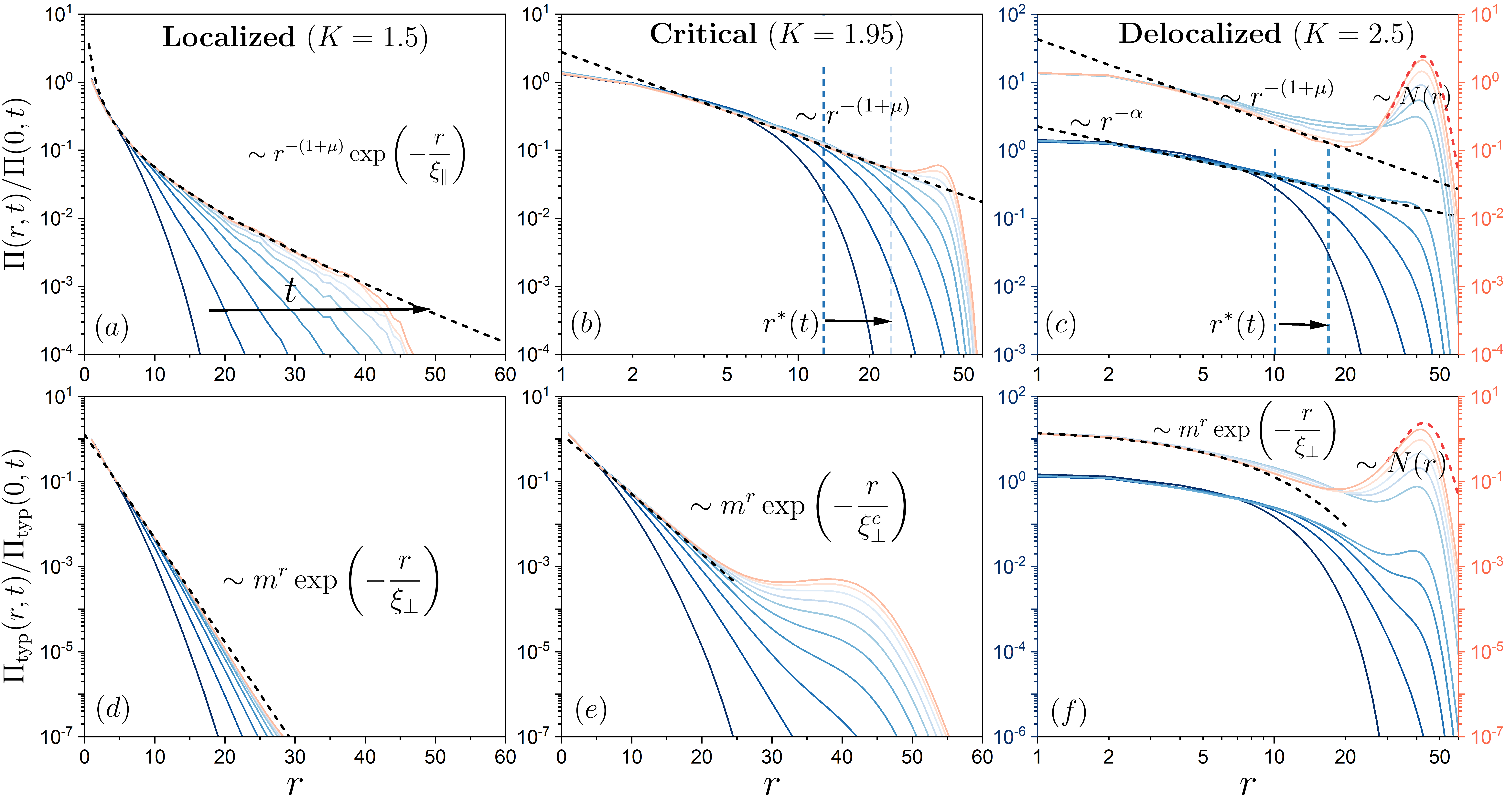}%

\caption{\label{FigWavePacket} Wave packet (WP) dynamics across the Anderson transition on Small-World graphs: The upper panels (a–c) show the average WP profile $\Pi(r,t)$, while the lower panels (d–f) display the typical WP profile $\Pi_{\mathrm{typ}}(r,t)$. Curves from dark blue to pale orange correspond to evolution times $t=18,48,126,329,853,2212,5736,14873,38566,10^5$.  
A strong distinction between average and typical WPs emerges in the localized and critical regimes and at times shorter than the ergodic time in the delocalized regime (lower blue curves in (c) and (f)), highlighting non-ergodicity. Notably, the typical WP saturates after the short typical localization time $\tau_{\mathrm{loc}}^{\mathrm{typ}}$ (see also Fig.~\ref{Fig2}(b), inset), forming an exponentially localized profile with localization length $\xi_\perp$, even at criticality (where $\xi_\perp^c\approx (2 \ln m)^{-1}$, a universal value \cite{PhysRevB.106.214202, PhysRevResearch.2.012020}). Conversely, the average WP exhibits slow logarithmic dynamics, a power-law spatial decay at criticality, and much slower convergence with exponential localization in the localized regime.  
Black dashed lines represent theoretical fits \cite{PhysRevB.99.214202, PhysRevResearch.6.L032024, PhysRevB.110.014210,PhysRevB.106.214202, PhysRevResearch.2.012020}. Blue dashed lines in (b) and (c) indicate the wavefront $r^*$, marking deviations from power-law decay \cite{PhysRevE.108.054127, PhysRevResearch.6.L032024, PhysRevB.110.014210}, with insets showing $r^* \sim \ln t$. Red dashed lines in (c) and (f) highlight the spatially ergodic behavior at large times and distances in the delocalized regime.  
In (c) and (f), the first five short-time curves are shifted vertically (blue $y$-axis), while the long-time curves align with the red $y$-axis for clarity. Results are averaged over 5400 disorder realizations with $N=2^{19}$ and $p=0.1$.  
}
\end{figure*}

The dynamics of wave packets have become a key observable in recent experiments with cold atoms and classical waves  \cite{PhysRevLett.103.155703,PhysRevLett.101.255702,PhysRevLett.108.095701}. These dynamics encode rich information about the critical properties of the Anderson transition \cite{PhysRevLett.101.255702,lemarie2009universality,OhtsukiKawarabayashi,PhysRevA.94.033615,PhysRevA.100.043612,PhysRevLett.124.186601,PhysRevE.108.054127,PhysRevResearch.6.L032024,PhysRevB.110.014210}.
Furthermore, in finite dimensions, kicked Floquet (time-periodic) systems, such as the kicked rotor \cite{santhanam2022quantum}, have proven exceptionally useful for studying Anderson localization. Notably, the experimental realization of the quasi-periodic kicked rotor has enabled the experimental determination of the critical exponent for the 3D and 4D Anderson transitions \cite{PhysRevLett.101.255702, PhysRevLett.108.095701, PhysRevLett.105.090601,PhysRevA.80.043626, madani2024exploring}. Numerically, the dynamics of such kicked models can be simulated with great efficiency using Fast Fourier Transforms, making it possible to reach the large times and system sizes that are crucial for studying critical dynamics.


In this letter, we address the critical dynamics of the $\text{AT}^{\infty}$  by introducing a unitary and kicked (Floquet) version of the Anderson model on SWGs \cite{watts1998collective,PhysRevLett.118.166801}—a variant of the Quantum Kicked Rotor (QKR) \cite{CHIRIKOV1979263,IZRAILEV1990299,PhysRevLett.75.4598,PhysRevA.29.1639}—which enables exact numerical simulations with large system sizes and long evolution times. 
We exploit the ability of our model to directly access the dynamics of the spatial profile of wave packets, allowing us
to study several critical properties of AT$^\infty$. 
Through this approach, we reveal the logarithmically slow dynamics in the non-ergodic critical regime, confirming recent random matrix predictions \cite{PhysRevResearch.6.L032024, PhysRevB.110.014210}. We contrast the distinct behaviors of 
typical and average wave packets across the localized phase up to criticality, confirming the existence of two localization lengths \cite{PhysRevResearch.2.012020, PhysRevB.106.214202} and revealing the existence of two localization times.
Using finite-time scaling analysis, we 
demonstrate the ergodic nature of the delocalized regime at large times and distances beyond non-ergodic bubbles. Finally we extract critical exponents and compare 
them with previous predictions and determinations from eigenstate and spectral properties \cite{PhysRevLett.118.166801, PhysRevB.99.024202,PhysRevB.99.214202, PhysRevB.110.184210}.


\emph{Unitary Anderson model on Small-World graphs (UASW).\textemdash} 
We construct the UASW starting from the 1D QKR \cite{CHIRIKOV1979263,IZRAILEV1990299,PhysRevLett.75.4598,PhysRevA.29.1639}, a paradigmatic model of quantum chaos that can be interpreted as a unitary Anderson model \cite{hamza2009dynamical}. The evolution operator for the QKR over a single period is defined as $
U_{1} = \sum_{i,j}^N e^{{\rm i}\Phi_i\delta_{ij}} \sum_{k=1}^{N} F_{ik} e^{-{\rm i}K V(2\pi k/N)} F_{kj}^{-1} |i\rangle\langle j|$,
where $V(x) = \cos x$, and $F_{jk} = e^{2{\rm i}\pi jk/N}/\sqrt{N}$ are discrete Fourier transform factors. Here, $i$ and $j$ are indices within $[\![1, N]\!]$, representing a 1D lattice of size $N$ with periodic boundary conditions. The phases $\Phi_i$ are randomly and uniformly distributed over $[0, 2\pi)$, and $K$ is the kicking strength, a crucial tuning parameter in this study that controls the hopping strengths between different sites.

To extend this unitary model to a SWG rather than a 1D lattice, we introduce long-range links with probability $0 < p < 1/2$ (the results presented here correspond to $p=0.1$). These links connect pairs of sites $\{j_\alpha, k_\alpha\}$, randomly chosen such that $|j_\alpha - k_\alpha| > 1$, resulting in $\lfloor pN \rfloor$ long-range links. This procedure constructs a SWG with an effective branching number $m \approx 1 + 2p$ \cite{watts1998collective}. 
The modification is implemented by multiplying the QKR evolution operator $U_1$ with unitary transition matrices $U_2^\alpha$ connecting $\{j_\alpha, k_\alpha\}$, $
U_{2}^\alpha = \frac{1}{\sqrt{2}} \left(|j_\alpha\rangle\langle j_\alpha| + |k_\alpha\rangle\langle k_\alpha| - |j_\alpha\rangle\langle k_\alpha| + |k_\alpha\rangle\langle j_\alpha|\right) 
+ \sum_{i \neq j_\alpha, k_\alpha}^N |i\rangle\langle i|$.
The successive application of these $\lfloor pN \rfloor$ unitary matrices can be expressed as $U_{2} = \prod_{\alpha=1}^{\lfloor pN \rfloor} U_{2}^\alpha$. The resulting $U = U_1 U_2$ defines the Floquet operator of our UASW.  

The kicked unitary nature of this model enables large-scale exact diagonalization through polynomial filtering techniques \cite{luitz2021polynomial}. This has allowed for a detailed finite-size scaling analysis of eigenstate and spectral properties of the UASW model, following \cite{PhysRevLett.118.166801,PhysRevResearch.2.012020,PhysRevB.106.214202}, to pinpoint the critical point at $K_c \approx 1.95$ for $p=0.1$ and characterize the critical properties of the transition, corroborating the findings of \cite{PhysRevLett.118.166801,PhysRevResearch.2.012020,PhysRevB.106.214202}. This analysis will be reported elsewhere \cite{LongPaper}.


\begin{figure}
\includegraphics[width=0.5\textwidth]{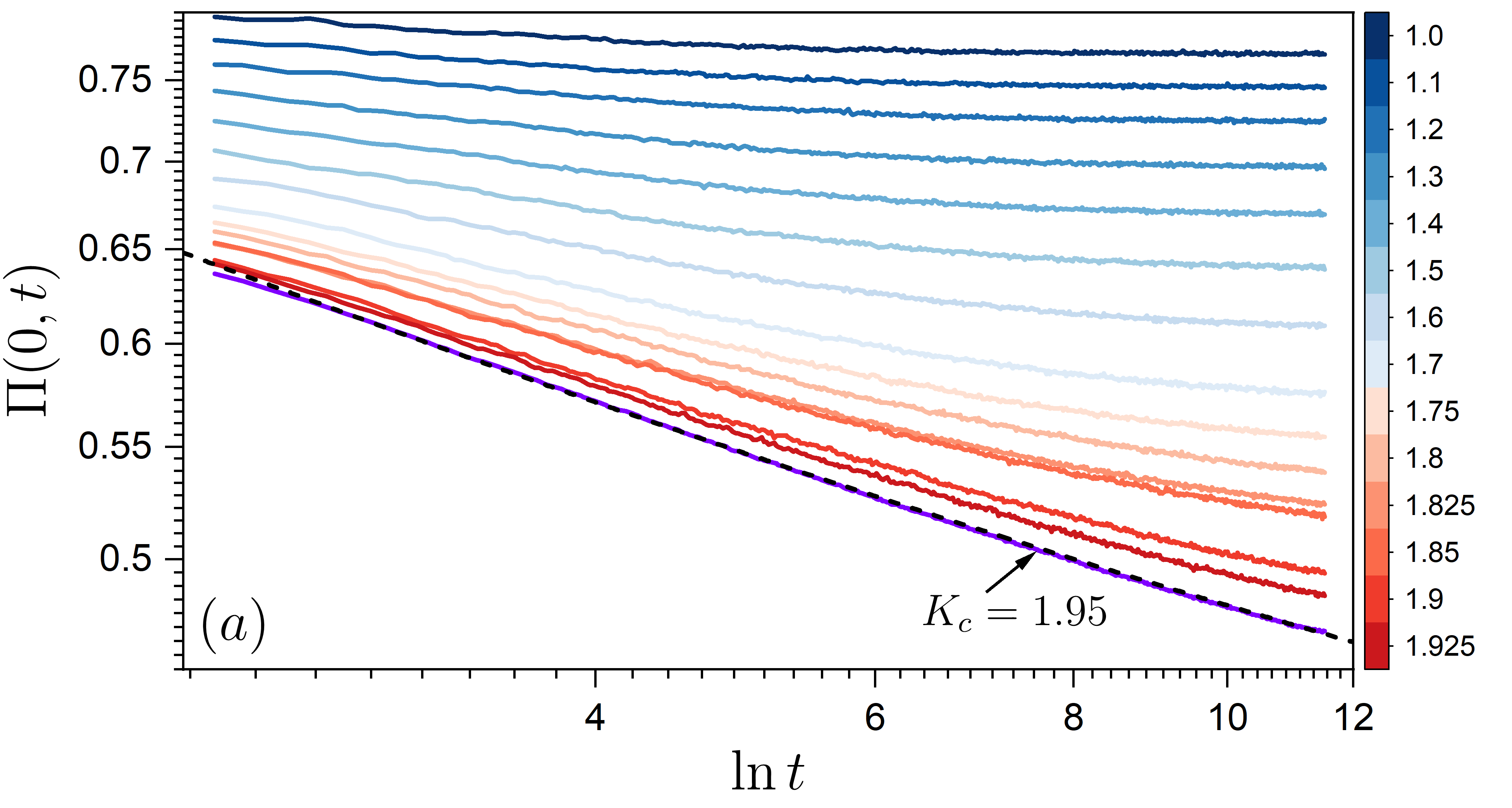}
\includegraphics[width=0.46\textwidth]{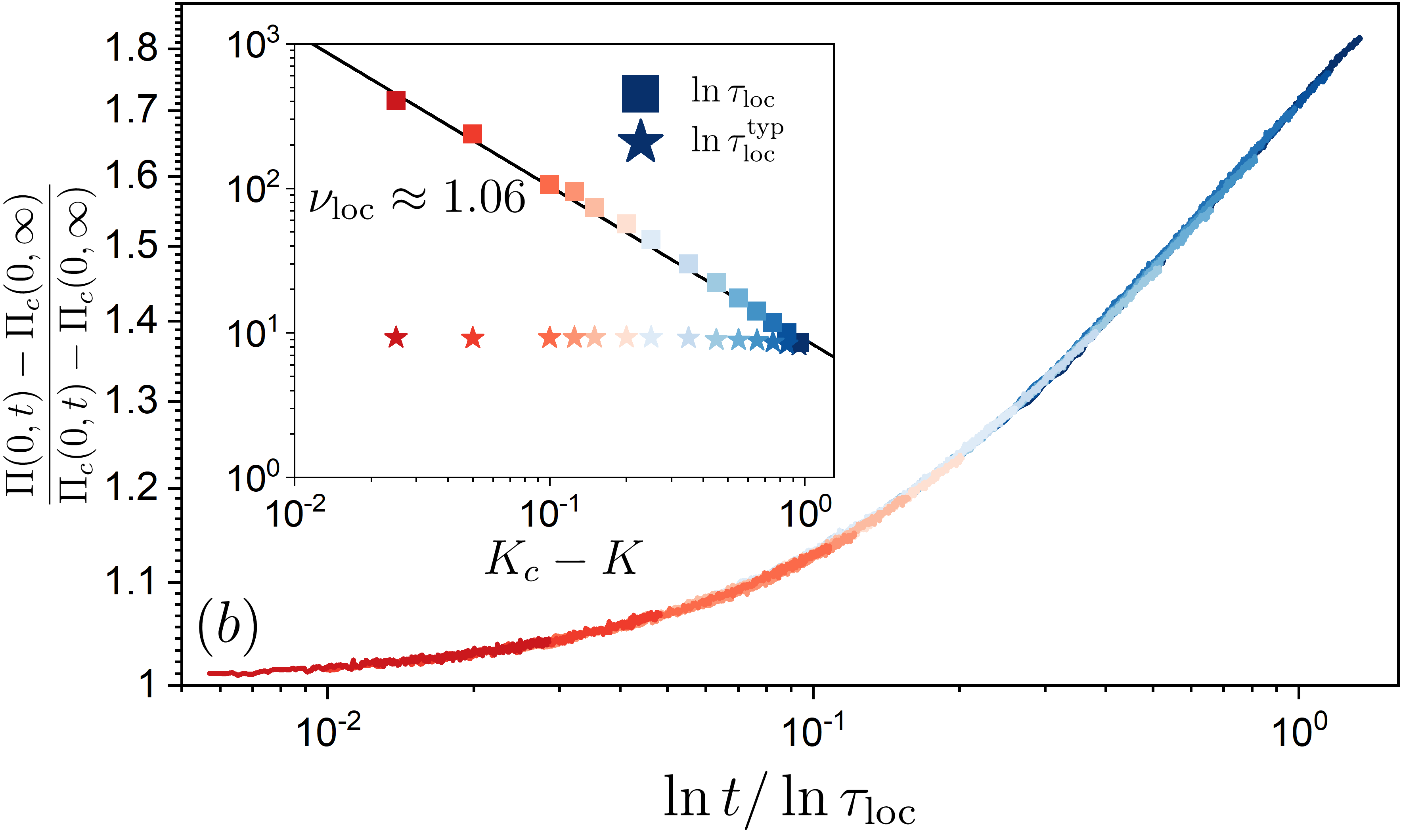}
\caption{\label{Fig2} Finite-time scaling of the return probability $\Pi(0,t)$ in the localized phase of the Anderson transition on Small-World graphs. $K$ varies from $K=1.0$ (strongest localization, upper blue) to $K=1.95$ (critical, lower purple curve), with evolution times ranging from $t=10$ to $t=10^5$. (a) Raw data for the coarse-grained $\Pi(0,t)$ (see text) show a slow convergence to a finite value, indicating asymptotic localization. At the critical point $K_c=1.95$, the return probability $\Pi_c(0,t)$ decays logarithmically as $\Pi_c(0,t) \approx A(\ln t)^{-\mu}+\Pi_c(0,\infty)$ (black dashed line), with $\mu=0.24$ (fixed) and $A, \Pi_c(0,\infty) \approx 0.107$ as fitting parameters, signaling critical localization \cite{PhysRevResearch.6.L032024, PhysRevB.110.014210,PhysRevB.99.214202}. (b) Single-parameter scaling of $\Pi(0,t)$. The data in (a) collapse onto a single scaling function when plotted as $\ln t / \ln \tau_{\mathrm{loc}}$, where $\tau_{\mathrm{loc}}$ is the average localization time. The inset shows $\ln \tau_{\mathrm{loc}}$ vs. $K$, along with the typical localization time $\tau_{\mathrm{loc}}^{\mathrm{typ}}$, defined as the time beyond which $\Pi_{\mathrm{typ}}(r=20,t)$ saturates (see Fig.~\ref{FigWavePacket}(d,e)). While $\tau_{\mathrm{loc}}^{\mathrm{typ}}$ saturates at a finite value at the transition, the average localization time logarithm diverges algebraically as $\ln \tau_{\mathrm{loc}} \sim |K - K_c|^{-\nu_{\text{loc}}}$ with $\nu_{\mathrm{loc}} \approx 1.06$, in perfect agreement with the critical exponent $\nu=1$ for averaged observables in the localized phase \cite{PhysRevLett.118.166801, PhysRevResearch.2.012020, PhysRevB.106.214202}. Results are averaged over $3,600$ to $18,000$ disorder configurations for a system size $N = 2^{17}$ and for $p=0.1$.  
 } 

\end{figure}



\emph{{WP dynamics across the Anderson transition in the UASW model.}\textemdash} 
Here we discuss the time evolution of wave packets (WP) which provides a direct probe of the critical dynamical properties. For a \jm{wave function} initially localized \jm{on} a single site, \( |\psi(t=0)\rangle = |i_0\rangle \) we define the \textit{average} and \textit{typical} WP as 
$\Pi(r,t) = \left\langle \sum_{i:d(i,i_0)=r} |\psi(i,t)|^2 \right\rangle$
and 
$\Pi_{\mathrm{typ}}(r,t) = \langle N(r)\rangle \exp\left\langle  \sum_{i: d(i,i_0)=r} \ln |\psi(i,t)|^2/N(r) \right\rangle$, respectively.
The distance \( r = d(i,i_0) \) is the graph (or Hamming) distance between site $i$ and the initial site $i_0$, and the sum runs over the $N(r)$ sites \( i \) at distance \( r \) from \( i_0 \), with \(\langle N(r) \rangle \sim m^r \) (\( m\approx1.3 \) for $p=0.1$).

The distinction between the average and typical WPs, shown in Fig.~\ref{FigWavePacket}, is made to shed light on the nonergodic properties of the localized (\( K=1.5 \), panels $(a,d)$) and critical (\( K=1.95 \), panels $(b,e)$) regimes, and at times shorter than the ergodic time in the delocalized regime (\( K=2.5 \), panels $(c,f)$) \cite{PhysRevResearch.2.012020, PhysRevB.106.214202}. The different behaviors of these WPs are discussed in the next sections.



\emph{{Logarithmically slow critical dynamics.}\textemdash} 
At the critical point, the spatial profile decays algebraically
$\Pi(r,t) \sim \Pi(0,t) r^{-(1+\mu)}$ for $r < r^*(t)$ and decays rapidly for $r > r^*(t)$, where $r^*(t)$ characterizes the wavefront. 
From the numerical data shown in Fig.~\ref{FigWavePacket} (b), we find $\mu = 0.24$ and $r^*(t) \sim \ln t$. These dynamical properties of the WP reproduce the \textit{critical localization} behavior predicted by random-matrix models of $\text{AT}^\infty$ \cite{PhysRevResearch.6.L032024, PhysRevB.110.014210}, similar to the behavior predicted for RRG (with $\mu = 0.5$ in that case) \cite{PhysRevB.99.214202}. The spatial profile of the typical WP shown in Fig.~\ref{FigWavePacket} (e) is distinct from the average one, and converges after a finite typical localization time $\tau_{\mathrm{loc}}^{\mathrm{typ}}$ (see inset of Fig.~\ref{Fig2}(b)) to a localized profile with the finite localization length $\xi_{\perp}^c \approx (2\ln m)^{-1}$, i.e., $\Pi_{\mathrm{typ}} (r,t)/\Pi_{\mathrm{typ}} (0,t) \sim m^r e^{-r/\xi_{\perp}^c}$, consistent with Refs.~\cite{PhysRevResearch.2.012020, PhysRevB.106.214202}.

The critical behavior is also encoded in the dynamics of the return probability, $\Pi(0,t)$, which is predicted to decay as $\Pi_c(0,t) \sim (\ln t)^{-\mu} + \Pi_c(0,\infty)$ \cite{PhysRevResearch.6.L032024, PhysRevResearch.2.012020, PhysRevB.110.014210, PhysRevB.99.024202}. This logarithmically slow decay is extremely challenging to characterize numerically. To mitigate fluctuations and considering the relationship between $\Pi(0,t)$ and the inverse participation ratio \( P_2 \) at large times, it is important to work with coarse-grained wave packets (WPs), starting from an initial flat distribution over \( N_{\text{init}} > 1 \) sites. In Fig.~\ref{Fig2}(a), we show the return probability of coarse-grained WPs initialized as 
\(
|\psi(t=0)\rangle = \sum_{i_0=1}^{N_{\text{init}}} \frac{e^{i\phi_{i_0}}}{\sqrt{N_{\text{init}}}} |i_0\rangle,
\)
where \( \phi_{i_0} \) are independent random phases that we average over, and \( N_{\text{init}} = 32 \).
The lowest curve represents $\Pi_c(0,t)$ for system size $N = 2^{17}$ and large evolution times up to $t = 10^5$ and is found compatible with the predicted behavior with $\mu = 0.24$ (determined from Fig.~\ref{FigWavePacket} (b)) and a small fitted value $\Pi_c(0,\infty) \approx 0.1$. 

\emph{Localized dynamics and finite-time scaling hypothesis.\textemdash} The localized phase exhibits significant non-ergodic properties \cite{PhysRevResearch.2.012020, PhysRevB.106.214202}, as highlighted by the distinct spatial profiles of the average and typical WPs (see Fig.~\ref{FigWavePacket} (a) and (d)). The average WP converges slowly, after an average localization time $\tau_{\mathrm{loc}}$ which diverges at the transition (see Fig.~\ref{Fig2}), into a profile characterized by the critical WP modulated by exponential decay, governed by the average localization length $\xi_\parallel$, i.e., $\Pi (r,\infty)/\Pi (0,\infty)\sim r^{-(1+\mu)}e^{-r/\xi_\parallel}$. On the other hand, the typical WP rapidly adopts, after a small typical $\tau_{\mathrm{loc}}^{\mathrm{typ}}$ that saturates at a finite value at the transition, a purely exponential localized profile with decay controlled by $\xi_{\perp}\ll \xi_\parallel$, specifically, $\Pi_{\mathrm{typ}} (r,\infty)/\Pi_{\mathrm{typ}} (0,\infty)\sim m^r e^{-r/\xi_{\perp}}$. These observations strongly corroborate the existence of two distinct localization lengths \cite{PhysRevResearch.2.012020, PhysRevB.106.214202, PhysRevResearch.6.L032024, PhysRevB.110.014210, zirnbauer2023wegner, vanoni2023renormalization, LongPaper} but also reveal the existence of two localization times.

To characterize the divergence of $\tau_{\text{loc}}$ near criticality, we perform a finite-time scaling analysis of the return probability $\Pi(0,t)$. In the infinite-time limit, $\Pi(0,\infty) \simeq P_2 \approx  \xi_{\parallel}^{-\mu} + P_2^{\infty}$ \cite{PhysRevB.103.064204, PhysRevE.108.054127, LongPaper}. Drawing an analogy with the finite-size scaling of eigenstate moments, we propose the following finite-time scaling behavior for the dynamics \cite{PhysRevLett.105.090601,PhysRevA.80.043626,madani2024exploring,PhysRevLett.118.166801,PhysRevB.106.214202,LongPaper}: 
\begin{equation}
    \frac{\Pi(0,t) - \Pi_c(0,\infty)}{\Pi_c(0,t) - \Pi_c(0,\infty)} = F\left(\frac{\ln t}{\ln \tau_{\text{loc}}}\right), 
\end{equation}
with $F(X) \rightarrow 1$ for $X \ll 1$ and $F(X) \rightarrow c X^{\mu}$ for $X \gg 1$. This scaling law in $\ln t/\ln \tau_{\text{loc}}$ is analogous to the linear finite-size scaling in $\ln N/\xi_{\parallel}$ found in \cite{PhysRevResearch.2.012020,PhysRevB.106.214202,PhysRevB.99.024202,LongPaper} as times scale like volumes. In Fig.~\ref{Fig2}(b), we present the collapse of numerical data when plotted as $\ln t/\ln \tau_{\text{loc}}$, revealing a divergence behavior of $\ln \tau_{\text{loc}} \sim |K - K_c|^{-\nu_{\text{loc}}}$ with $\nu_{\text{loc}} \approx 1$, in perfect agreement with the critical exponent $\nu=1$ for averaged observables in the localized phase \cite{PhysRevLett.118.166801, PhysRevResearch.2.012020, PhysRevB.103.064204}. 

\begin{figure}

\includegraphics[width=0.49\textwidth]{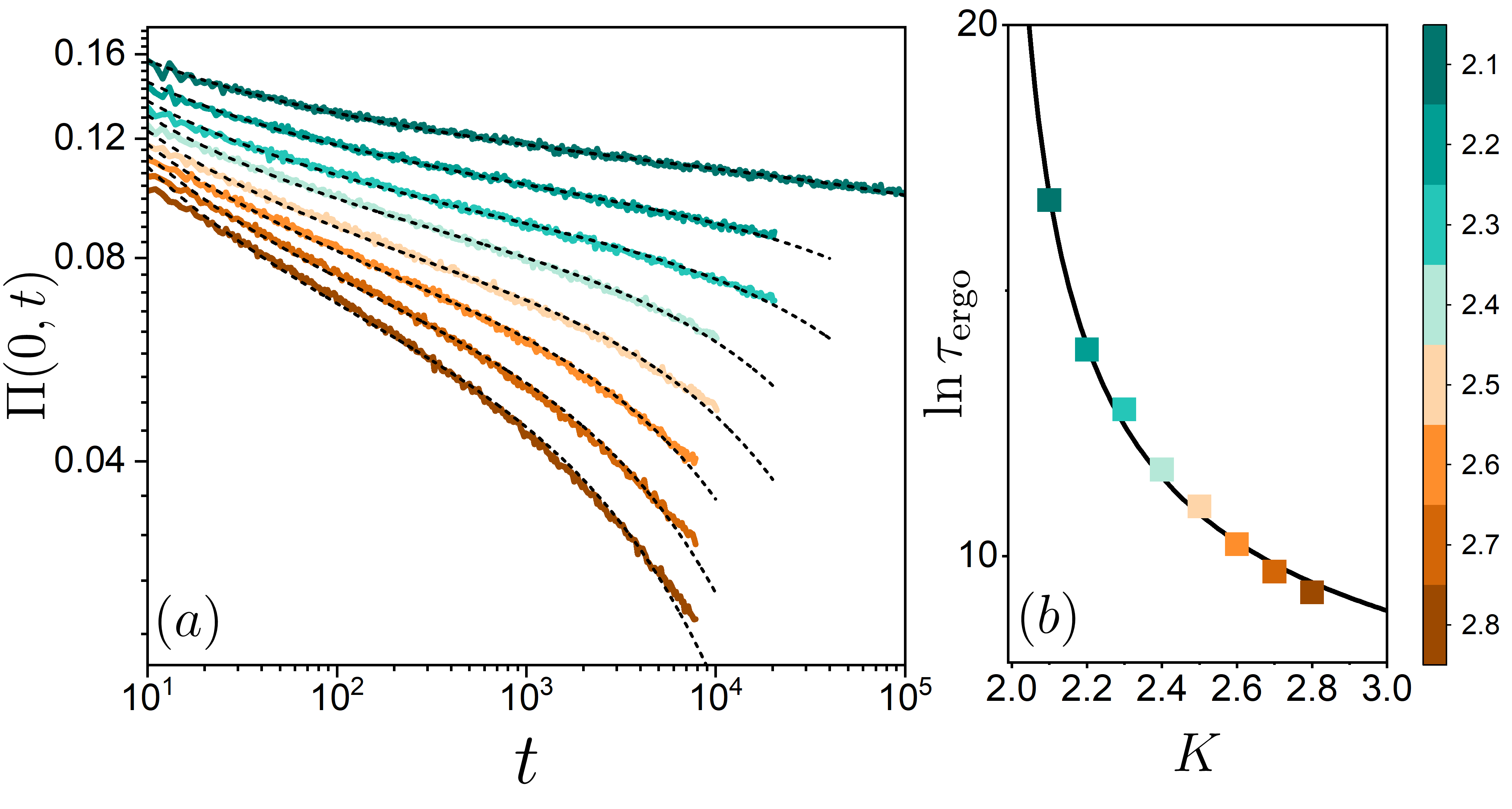}

\caption{\label{FigDelocalized} Ergodic time scaling in the delocalized phase.
(a) Time evolution of the return probability $\Pi(0,t)$ for $K=2.1, 2.2, ..., 2.8$ (top to bottom). To minimize finite-size effects, data beyond the point where dynamics for system sizes $N = 2^{17}$ and $N = 2^{18}$ deviate are excluded. The dashed lines represent fits by $A(\ln t)^{-d}\exp\left[-(t/\tau_{\mathrm{ergo}})^{\beta_d}\right]$ with $\beta_d=0.61$ the fixed stretch exponent of classical diffusion on SWGs we determined independently, and $A, d, \tau_{\mathrm{ergo}}$ as fitting parameters. Results are averaged over 18,000 disorder configurations.
(b) Dependence of the ergodic time $\tau_{\mathrm{ergo}}$ on $K>K_c$, extracted from the fits. The solid line shows a fit of the form $A|K-K_c|^{-\nu}+C$, with $\nu=0.5$ (fixed).}

\end{figure}

\emph{Delocalized dynamics and ergodicity.\textemdash} 
In the delocalized regime, the average WP profile, shown in Fig.~\ref{FigWavePacket}(c), exhibits two different behaviors as a function of time. (I) At sufficiently small times, the profile follows a power-law spatial decay $\Pi(r, t) / \Pi(0, t) \sim r^{-\alpha(K)}$ with $\alpha(K) < \mu$, resembling the critical behavior but with a different exponent. When $\alpha(K) < 1$, this indicates \textit{logarithmic multifractality}, as introduced in random-matrix ensembles \cite{PhysRevResearch.6.L032024, PhysRevB.110.014210}. A wavefront $r^*(t)$, beyond which the WP decays rapidly, follows $r^*(t) \sim \ln t$. (II) At large times, a non-ergodic radius emerges. The WP profile stabilizes to $\Pi(r, t) / \Pi(0, t) \sim r^{-\gamma}$ with $\gamma \approx 1.5$ at sufficiently small distances, but crosses over to ergodic behavior $\Pi(r, t) / \Pi(0, t) \sim N(r)$ at large distances. Thus, beyond a characteristic time or spatial scale, ergodicity is restored. This is further supported by the typical WP in Fig.~\ref{FigWavePacket}(f), where ergodic behavior reappears at long times and large distances. However, the average and typical WPs differ at short times and distances, revealing non-ergodic bubbles.

To quantitatively assess the ergodic time $\tau_{\mathrm{ergo}}$, which separates non-ergodic critical-like behavior from ergodic transport, we analyze the return probability $\Pi(0,t)$. Care is needed in its interpretation. First, delocalized dynamics suffers from strong finite-size effects, in addition to the finite-time effects we focus on. To minimize these, we exclude data beyond the time when the results for system sizes $N = 2^{17}$ and $N = 2^{18}$ start to diverge from each other. Second, classical diffusion on SWG exhibits ``glassy'' relaxation, with the return probability decaying as a stretched exponential at long times, $\Pi(0,t) \sim t^{-1/2}e^{-ct^{\beta_d}}$, where $c$ depends on $m$ \cite{PhysRevB.38.11461, monasson1999diffusion, PhysRevE.62.4405}. Our simulations of classical diffusion on SWG yield $\beta_d \approx 0.61$ (see \cite{LongPaper}).  This contrasts with the exponential decay found in RRG \cite{PhysRevB.99.024202} and Bethe lattice \cite{Monthus1996}.


Based on the picture provided by WPs, at short times the system resides in a non-ergodic bubble, where the return probability decays logarithmically slowly, $\Pi(0,t) \sim (\ln t)^{-d(K)}$ \cite{PhysRevResearch.6.L032024, PhysRevB.110.014210}. We propose that the quantum decay follows the behavior:
\begin{equation}
\Pi(0,t) \sim (\ln t)^{-d(K)} e^{-[t/\tau_{\mathrm{ergo}}(K)]^{\beta_d}}\;.  
\end{equation} 
In Fig.~\ref{FigDelocalized} (a), the numerical data re well fitted by this formula, and the ergodic time shown in panel (b) aligns with the analytical prediction, $\tau_{\text{ergo}}(K) \sim e^{\mathrm{cst} |K-K_c|^{-\nu}}$, with $\nu = 0.5$ \cite{PhysRevB.99.024202}. Thus, $\tau_{\mathrm{ergo}}(K)$ behaves as the correlation volume \cite{PhysRevLett.118.166801, PhysRevB.99.024202,PhysRevB.99.214202, PhysRevB.110.184210}.


\emph{Conclusion.\textemdash}
We have characterized the critical dynamics of the Anderson transition in infinite dimensions using a unitary Anderson model on Small-World graphs, enabling large-scale, long-time simulations. By directly probing wave packet expansion, we reveal logarithmically slow dynamics at criticality, two distinct localization times in the localized phase, and ergodic diffusion at large scales and times in the delocalized regime. Finite-time scaling yields critical exponents $\nu_\text{loc} \approx 1$ for localization time and $\nu \approx 1/2$ for ergodic time. Our results confirm key analytical predictions \cite{PhysRevResearch.6.L032024, PhysRevB.110.014210, PhysRevB.99.214202, TIKHONOV2021168525} while challenging recent findings on other graphs and random matrices \cite{PhysRevB.98.134205, PhysRevB.101.100201, 10.21468/SciPostPhys.6.1.014, 10.21468/SciPostPhys.11.2.045}, highlighting the need for a systematic comparison of methods and models. Our study provides a quantitative framework for dynamical scaling that may offer insights into Many-Body Localization. Experimental realizations in cold-atom and quantum simulator platforms, where graph structures in Hilbert space can be engineered \cite{PhysRevLett.123.130601,yao2023observation}, appear within reach.

\acknowledgements
We wish to thank O. Giraud for
fruitful discussions. This study has been (partially) supported through the EUR grant NanoX ANR-17-EURE-
0009 in the framework of the ``Programme des Investisse-
ments d’Avenir", the France 2030 program (Grant No. ANR-23-PETQ-0002), the French-Argentinian IRP COQSYS, research funding Grants No.~ANR-18-CE30-0017, ANR-19-CE30-0013 and ManyBodyNet, and by the Singapore Ministry of Education Academic Research Funds Tier II (WBS No. A-8001527-02-00 and A-8002396-00-00). We thank Calcul en Midi-Pyrénées
(CALMIP), the National Supercomputing Centre (NSCC) of Singapore, and the Consortium des Equipements de Calcul Intensif (CECI), funded by the Fonds de la Recherche Scientifique de Belgique (F.R.S.-FNRS) under Grant No. 2.5020.11 for computational resources and assistance. W.~Chen is supported by the President's Graduate Fellowship at National University of Singapore and the Merlion Ph.D. Scholarship awarded by the French Embassy in Singapore. I.G.-M. received support from Argentinian Agencia I$+$D$+$i (Grants No. PICT-2020-SERIEA-00740 and PICT-2020-SERIEA-01082) as well as  from CNRS (France) through the International Research Project (IRP) “Complex Quantum Systems” (CoQSys).

\bibliography{refs.bib}

\end{document}